# Observation of Topologically Protected States at Crystalline Phase Boundaries in Single-layer WSe$_2$

Miguel M. Ugeda[1,2,3*], Artem Pulkin[4], Shujie Tang[5,6], Hyejin Ryu[5,7], Quansheng Wu[4,8], Yi Zhang[5,6,9], Dillon Wong[10], Zahra Pedramrazi[10], Ana Martín-Recio[10,11], Yi Chen[10], Feng Wang[10,12,13], Zhi-Xun Shen[6,14], Sung-Kwan Mo[5], Oleg V. Yazyev[4,8] and Michael F. Crommie[10,12,13*]

[1]*Centro de Física de Materiales (CSIC-UPV-EHU), Manuel Lardizábal 5, 20018 San Sebastián, Spain.*

[2]*Ikerbasque, Basque Foundation for Science, 48013 Bilbao, Spain.*

[3]*Donostia International Physics Center (DIPC), 20018 San Sebastián, Spain.*

[4]*Institute of Physics, Ecole Polytechnique Fédérale de Lausanne (EPFL), CH-1015 Lausanne, Switzerland.*

[5]*Advanced Light Source, Lawrence Berkeley National Laboratory, Berkeley, California 94720, USA.*

[6]*Stanford Institute for Materials and Energy Sciences, SLAC National Accelerator Laboratory, Menlo Park, California 94025, USA.*

[7]*Center for Spintronics, Korea Institute of Science and Technology, Seoul 02792, Korea.*

[8]*National Centre for Computational Design and Discovery of Novel Materials MARVEL, Ecole Polytechnique Fédérale de Lausanne (EPFL), CH-1015 Lausanne, Switzerland.*

[9]*National Laboratory of Solid State Microstructures, School of Physics, Collaborative Innovation Center of Advanced Microstructures, Nanjing University, Nanjing 210093, China.*

[10]*Department of Physics, University of California at Berkeley, Berkeley, California 94720, USA.*

[11]*Departamento de Física de la Materia Condensada, Universidad Autónoma de Madrid, E-28049 Madrid, Spain.*

[12]*Materials Sciences Division, Lawrence Berkeley National Laboratory, Berkeley, California 94720, United States.*

[13]*Kavli Energy NanoScience Institute at the University of California Berkeley and the Lawrence Berkeley National Laboratory, Berkeley, California 94720, United States.*

[14]*Geballe Laboratory for Advanced Materials, Departments of Physics and Applied Physics, Stanford University, Stanford, California 94305, United States.*

*\* Corresponding authors: mmugeda@dipc.org and crommie@berkeley.edu.*

**Transition metal dichalcogenide (TMD) materials are unique in the wide variety of structural and electronic phases they exhibit in the two-dimensional (2D) single-layer limit. Here we show how such polymorphic flexibility can be used to achieve topological states at highly ordered phase boundaries in a new quantum spin Hall insulator (QSHI), 1T'-WSe$_2$. We observe helical states at the crystallographically-aligned interface between quantum a spin Hall insulating domain of 1T'-WSe$_2$ and a semiconducting domain of 1H-WSe$_2$ in contiguous single layers grown using molecular beam epitaxy (MBE). The QSHI nature of single-layer 1T'-WSe$_2$ was verified using ARPES to determine band inversion around a 120 meV energy gap, as well as STM spectroscopy to directly image helical edge-state formation. Using this new edge-state geometry we are able to directly confirm the predicted penetration depth of a helical interface state into the 2D bulk of a QSHI for a well-specified crystallographic direction. The clean, well-ordered topological/trivial interfaces observed here create new opportunities for testing predictions of the microscopic behavior of topologically protected boundary states without the complication of structural disorder.**

Materials exhibiting the quantum spin Hall effect (QSHE) create new opportunities for directly imaging the spatial extent of topologically protected 1D edge states and for determining how they interact with bulk states and defects. Such systems, however, can be difficult to isolate and to access via microscopy. HgTe and InAs/GaAs quantum wells, for example, are well known QSHIs[1,2], but are not easily accessible to high resolution scanned probe microscopy because they are buried interface systems. Bi-based surface systems (predicted to be QSHIs[3,4]) are more accessible to scanned probe microscopy, but exhibit strong substrate interactions[5–7]. Monolayer TMD materials (MX$_2$ where M = Mo, W, and X = S, Se, Te) in the distorted octahedral 1T' phase, on the other hand, are a new class of QSHIs[8] that retain their topological properties on different substrates and are completely accessible to high resolution scanned probe microscopy[9–11]. Monolayer 1T'-WTe$_2$ films have recently been shown to exhibit all of the hallmarks of the QSH effect (e.g., band inversion, helical edge-states, and edge-state quantum conduction) via angle-resolved photoemission spectroscopy (ARPES)[9], scanning tunneling microscopy/spectroscopy (STM/STS),[9–11] and transport measurements[12,13]. Monolayer 1T'-WTe$_2$, however, poses challenges for quantitative microscopy of topological edge states due to the high degree of structural disorder in the edges of 2D 1T'-WTe$_2$ islands produced by MBE. Although the existence of topological edge states is protected against disorder, quantitative characterization of their decay lengths, dispersion features, and defect interactions requires crystallographically well-ordered edges.

In order to achieve structurally well-defined boundaries in a fully accessible QSHI, we grew mixed-phase WSe$_2$ monolayers on SiC(0001) using MBE growth techniques. Single-layer WSe$_2$ is bimorphic with two stable crystalline phases (1H and 1T' (Fig. 1a)) that are close in energy[8], thus enabling the growth of mixed topological/trivial phases with crystallographically defined phase boundary interfaces. The 1H phase (which is the structural ground state of WSe$_2$) has a much larger electronic bandgap[14,15] than the 1T' phase, thus allowing the two phases to be easily distinguished. The onset of the QSHE in mixed-phase WSe$_2$ thus results in topologically protected states at crystallographically well-defined 1T'-1H phase boundary interfaces. We are able to verify the QSHI ground state of 1T'-WSe$_2$ using ARPES, STM/STS, and first-principles calculations. ARPES reveals the existence of inverted bands at E$_F$ and the presence of a bulk bandgap. STS measurements confirm the bulk bandgap seen by ARPES and further demonstrate the existence of topological interface states within this bandgap and that are spatially localized at 1T'-WSe$_2$ boundaries. These boundary states are easily observable at crystallographically well-ordered 1T'-1H interfaces, but can also be seen at irregular 1T' edges. The structural perfection of the 1T'-1H boundary allows us to measure an interface state decay length of 2 nm into bulk 1T'-WSe$_2$, which agrees with the results of *ab initio* numerical simulations.

Our experiments were carried out on high-quality single-layers of WSe$_2$ grown on epitaxial bilayer graphene (BLG) on 6H-SiC(0001) by MBE. In order to obtain the metastable 1T'-WSe$_2$ phase, the temperature of the BLG/SiC(0001) substrate was held at 500 K during growth, a significantly lower temperature than required to grow the more stable 1H phase (675 K). Under these growth conditions the RHEED pattern of single-layer WSe$_2$ (Fig. 1b) shows the formation of an additional large lattice periodicity (5.8 Å) consistent with the 1T' phase that coexists with the shorter 1H phase periodicity (3.3 Å). XPS measurements of the WSe$_2$ layers (Fig. 1c) reveal the emergence of two new pairs of peaks ($d^{T'}$ and $f^{T'}$) near the characteristic Se ($d^H$) and W ($f^H$) peaks for the 1H phase[15], suggesting the presence of an additional lattice symmetry for W and Se[16]. STM imaging confirms that our WSe$_2$ layers are composed of coexisting domains of 1H and 1T' phase (Figs. 3a, 5a and SI). Figure 1d shows an atomically-resolved STM image of the 1T' phase of WSe$_2$, which is characterized by straight atomic rows of two non-equivalent zigzag atomic chains. The 1T' phase of Fig. 1d exhibits a period-enlargement to 5.73 ± 0.09 Å along the *x* direction compared to the 1H phase, in good agreement with the RHEED spectra. Adjacent atomic rows in 1T'-WSe$_2$ exhibit a slight translational shift along the *y*-direction due to a shear angle that varies between 2° and 6° depending on the domain, similar to that observed previously for other TMD materials[17,18]. We identify the atomic rows in the STM images of Figs. 1d,e as originating from W-Se zigzag chains (see sketch in Fig. 1e), in good agreement with the expected structural distortion of the 1T' phase[8]. The ball-and-stick model shown in Figs. 1a,e corresponds to our calculated relaxed atomic structure of 1T'-WSe$_2$.

We experimentally characterized the electronic structure of coexisting 1H and 1T' phases of single-layer WSe$_2$ via ARPES and STS. Figure 2c shows the Fermi surface (FS) intensity map for a 0.8 monolayer (ML) coverage of mixed-phase WSe$_2$ measured via ARPES. The observed FS structure is entirely due to the 1T' phase since the valence band (VB) maximum of 1H-WSe$_2$ has a much higher binding energy at E = -1.1 eV[15]. The FS is composed of two small elliptical electron pockets at the Λ points located along ΓY (Fig. 2a). The three equivalent rotational domains of the 1T' phase on BLG leads to the emergence of three pairs of these features rotated by 120° (Fig. 2b), thus forming a ring-like FS around the Γ point. Figure 2d shows the measured band dispersion along the ΓY direction of the Brillouin zone (BZ). Due to the rotational domains, contributions from both the ΓY and ΓP directions can be resolved. The VB maximum is approximately 170 ± 20 meV below the Fermi energy (E$_F$) and exhibits a flattened, non-parabolic onset shape along ΓY. Naturally occurring n-type doping in our samples shifts the conduction band (CB) below the Fermi energy, which is why the electron pockets at Λ are visible in the ARPES spectrum. This reveals the existence of an indirect bandgap (E$_g$) that can be quantified by taking the difference of the energy positions of the CB minimum (at the Λ point) and the VB maximum (at the Γ point) from two energy distribution curves (EDCs) of the ARPES spectrum (taken along the dashed lines in Fig. 2d). As shown in Fig. 2f, we extract a bandgap value of E$_g$ = 120 ± 20 meV centered at E = -110 meV ± 20 meV. The observed band dispersion and gap value is characteristic of band inversions predicted for 1T'-TMD materials[8].

The local density of states (LDOS) of mixed-phase, single-layer WSe$_2$ was measured via STS point spectroscopy, as seen in Fig. 3a. The 1H phase of monolayer WSe$_2$ shows a bandgap of 1.94 eV, in good agreement with previous measurements[15], but the 1T' phase reveals a finite, asymmetric LDOS that extends across both the occupied state and unoccupied state regions. The most pronounced feature in the unoccupied state region of the 1T' phase is a broad, asymmetric peak centered around + 0.24 V. The finite LDOS seen in the occupied state region of the 1T' phase (-1 V < $V_s$ < 0 V) confirms that the bands observed in ARPES at low binding energy (Fig. 2d) belong to the 1T' phase since this energy range is clearly gapped out for the 1H phase. Also prominent in the electronic structure of the 1T' phase is a gap-like feature located at $V_s$ = -130 ± 5 mV. Fig. 3c shows a close-up of this feature (the boxed region of Fig. 3a). The width of this 1T' gap feature can vary depending on surface position, but it has an average FWHM = 85 mV ± 21 meV (see SI for gap statistics). A second dip feature located at $E_F$ can be seen in the dI/dV curves taken for 1T'-WSe$_2$. A similar zero-bias feature has also been seen in 1T'-WTe$_2$ and has been attributed to the opening of a Coulomb gap[19]. These characteristic features are seen throughout the 1T' bulk region for islands with the narrowest widths larger than ~8 nm. For 1T' islands of smaller widths the zero-bias feature is replaced by a larger size-dependent energy gap that opens at $E_F$ and the 1T' gap feature vanishes, ostensibly due to size quantization effects[20]. The bulk gap feature observed by STM spectroscopy at $V_S$ = -130 mV is consistent with the ARPES bulk bandgap for 1T'-WSe$_2$ when lifetime broadening effects are taken into account (see SI). Such broadening likely arises from a combination of electronic, vibrational, and defect based scattering, as well as coupling to the graphene substrate[21].

In order to further understand the electronic structure of single-layer 1T'-WSe$_2$, we also characterized its quasiparticle interference (QPI) patterns near $E_F$ via Fourier transform (FFT) analysis of dI/dV images. Figures 4b-d show constant-bias dI/dV maps taken in the same pristine region of 1T'-WSe$_2$ for energies within the CB (b and c) as well as in the VB (d). The QPI patterns observed in the dI/dV maps exhibit long-range oscillations with wave fronts parallel to the *x*-direction and closely spaced rows aligned parallel to the *y*-direction (i.e., the atomic rows). The corresponding FFT images of the conductance maps (Figs. 4e-h) show distinct features that reflect the band structure contours at these different energies.

The electronic features we have described up to now for bulk single-layer 1T'-WSe$_2$ are consistent with an inverted bandgap and the QSHI phase. A key feature of QSHIs, however, is the existence of helical states at the boundaries. WSe$_2$ is particularly well-suited to explore the existence of such states due to the coexistence of the 1T' and 1H phases, which leads to straight, defect-free interfaces as shown in Figs. 3a and 5a. Fig. 5b shows a color-coded series of dI/dV spectra measured along the 5.3 nm-long black arrow in Fig. 5a oriented perpendicular to the 1T'-1H interface (the interface is marked by a dashed white line). The 1T'-1H interface is defined as the point where the STM topograph height reaches 50% of the height difference from the 1H average terrace height to the 1T' average terrace height for $V_s = -0.52$ V, $I = 0.2$ nA. This definition is also valid for other biases within the range $-0.6$ V $< V_s < -0.1$ V and $I_t \leq 0.5$ nA (the 1T' terrace is $2.9 \pm 0.2$ Å higher than the 1H terrace under these standard tunneling conditions). Fig. 5b shows that the STS feature identified as the bulk bandgap at -130 meV is present in the bulk 1T' material only for distances greater than 2 nm from the 1T'-1H interface.

The 1T'-WSe$_2$ bulk gap disappears at distances closer than 2 nm from the 1T'-1H interface and a prominent peak emerges in the LDOS at the same energy that previously showed a gap. This is illustrated in Fig. 5c which shows dI/dV curves taken in the bulk region (orange curve) and in the edge region (blue curve) as indicated by the dashed lines in Fig. 5b. The emergence of this peak is consistent with the existence of a 1D conducting helical state as expected in a QSHI. In order to resolve the spatial extent of the interface state, we mapped the dI/dV conductance near the 1T'-1H interface with sub-nm resolution. Fig. 5d shows a dI/dV map of the same region shown in Fig. 5a at the bias voltage at the center of the interface-state peak ($V_s$ = -130 meV). This map shows bright intensity in the 1H phase region near the 1T'-1H interface. This is due to electronic states from the 1T' phase "leaking into" the gapped 1H phase, similar to the phenomenon of metal-induced-gap-states (MIGS)[22]. Below the 1T'-1H interface in the 1T' phase region a very uniform band of increased dI/dV intensity can be seen that penetrates 2 nm into the 1T' bulk (marked "interface State"). This reveals the spatial extent of the topological interface state that resides in the bulk energy gap of single-layer 1T'-WSe$_2$ (see Fig. 5e for average linescan profile). The penetration depth of 2 nm that we extract from this linescan is in reasonable agreement with previous predictions for topological edge states.[8] (STM spectroscopy performed at the disordered edges of 1T'-WSe$_2$ islands also showed the spectral signatures of topologically protected edge-states, but in this case disorder prevented any quantitative determination of edge-state width (see SI)).

In order to better understand the topological behavior of this mixed phase system, we performed *ab initio* calculations using density functional theory (DFT) (see Methods). The resulting relaxed structure (Fig. 1a) is consistent with previous calculations for this phase[8] and agrees well with our STM topographic images (Fig. 1e). Figs. 2e and 4a show the band structure along Y-Γ-Y (red) and P-Γ-P (green) directions over a wide energy range calculated using a hybrid functional The results of our band structure calculations agree well with the ARPES results shown in Fig. 2 after performing a rigid shift of -130 meV to account for n-type doping observed in our samples. The non-parabolic flattened shape of the valence band near the Γ point unambiguously indicates the occurrence of band inversion, a prerequisite for topologically non-trivial behavior. The calculated band structure also shows an energy gap of 123 meV with band edges along the ΓY direction in reasonable agreement with both our ARPES and STS results.

Comparison of the calculated bulk 1T'-WSe$_2$ LDOS(E) with experimental STM dI/dV spectra shows qualitative agreement over a broad energy range as seen in Fig. 3b. The gap structure, the rise in valence band LDOS as energy is decreased, and the conduction band peak feature near 0.2 eV are all observed (although sharp LDOS features appear to be washed out in the data, likely due to lifetime broadening (see SI)). We also simulated 1T'-WSe$_2$ QPI patterns that take into account the band inversion and gap opening seen in Fig. 4a. Figs. 4i-k show the calculated QPI patterns for energies at + 100 meV, -40 meV and -300 meV in comparison to the experimental QPI patterns of Figs. 4e-h. Here the agreement is reasonable for features such as the multi-lobe structure along $k_y$ and the elongation along $k_x$, but remains qualitative overall due to limitations in the size of the 1T' phase domains that were imaged to obtain the experimental FFTs.

The calculated electronic structure for a single-layer WSe$_2$ 1T'-1H interface model structure is shown in Fig. 6. The proposed interface model (Fig. 6a) was chosen because its electronic structure best matches our experimental data. Although the experimental interface has a well-defined crystallographic orientation, it is not possible to verify its atomic structure due to limitations in experimentally resolving chemical bonds. The calculations of interface models have been performed using the standard DFT approach due to the large model size (see Methods). This results in a reduced band gap (29 meV) while all other band structure features are very similar to those resulting from the hybrid functional calculations. Fig. 6b shows the calculated dispersion of topologically-protected interface states running parallel to the 1T'-1H interface shown in Fig. 6a. A total of three bands span the bulk band gap. The odd number of bands is consistent with a topological origin and spin-momentum locking is clearly manifested. Fig. 6c demonstrates how extrema in the dispersion of these interface-state bands give rise to a large LDOS intensity within the bulk bandgap, consistent with the experimental dI/dV curve in Fig. 5c. The dependence of LDOS(E) on the distance from the 1T'-1H interface (Fig. 6d, black curve) shows that these states are localized within approximately 2 nm of the interface in the 1T' domain in reasonable agreement with the experimental interface-state decay length shown in Figs. 5d,e.

In conclusion, our measurement support the results of first-principles calculations and confirm the presence of the QSHI phase in single-layer 1T'-$WSe_2$. The ability to observe topologically protected interface-states at atomically smooth phase-boundary interfaces allows us to extract new quantitative information on these novel states, such as their penetration depth into the 1T'-$WSe_2$ bulk. This creates new opportunities for investigating topologically non-trivial electronic phases in 2D TMDs and takes us a step closer to the integration of 2D QSH layers into more complex heterostructures that exploit topologically protected charge and spin transport.

**Methods:**

Single-layer $WSe_2$ was grown by molecular beam epitaxy (MBE) on epitaxial BLG on 6H-SiC(0001) at the HERS endstation of Beamline 10.0.1, Advanced Light Source, Lawrence Berkeley National Laboratory (the MBE chamber had a base pressure of ~$3 \times 10^{-10}$ Torr). We used SiC wafers with resistivities of $\rho \sim 0.1$ Ω cm. The epitaxial BLG substrate was prepared by following the procedure detailed in ref. 14. High-purity W and Se were evaporated from an electron-beam evaporator and a standard Knudsen cell, respectively, while keeping the flux ratio of W to Se at 1:15. A Se capping layer with a thickness of ~10 nm was deposited on the sample surface after growth to protect the film from contamination and oxidation during transport through air to the ultrahigh vacuum scanning tunneling microscopy (UHV-STM) chamber. The Se capping layer was removed for STM experiments by annealing the sample to ~500 K in the UHV-STM system for 30 min. STM imaging and STS experiments were performed in an Omicron LTSTM operated at T = 4 K. STM differential conductance (dI/dV) spectra were measured using standard lock-in techniques. The STM tip was calibrated by measuring reference spectra on the graphene substrate in order to avoid tip artifacts. STM/STS data were analyzed and rendered using WSxM software [23].

First-principles calculations were performed using DFT within the generalized gradient approximation (GGA)[24] as implemented in the Quantum-ESPRESSO package[25] and within the HSE03[26] hybrid functional using the VASP package[27]. The single-particle Hamiltonian for valence and conduction states included relativistic corrections through ultrasoft pseudopotentials[28] adapted from Ref. [29]. The plane-wave basis set cutoff for wavefunctions was set to 80 Ry. Reciprocal space sampling was performed on an 11×18 k-point mesh in the rectangular Brillouin zone. The $WSe_2$ monolayers were decoupled along the out-of-plane direction by 1.5 nm of vacuum. Prior to calculating electronic properties, the atomic coordinates and in-plane lattice constants were fully relaxed. QPI patterns were calculated via the autocorrelation function of electronic bands as implemented in WannierTools[30]:

$$f(k,E) = \sum_{n,n'} \int \delta(E_n(k') - E)\delta(E_{n'}(k+k') - E)dk'$$

where $E_n(k)$ is the energy dispersion of the $n^{th}$ Bloch band. The autocorrelation functions presented in Fig. 4i-k were calculated on a fine 1200×2400 k-point mesh. We find that explicitly including the matrix elements does not qualitatively change the calculated QPI patterns. The electronic structure of a 1T'1H interface presented in Fig. 6 was calculated using the non-equilibrium Green's function technique[31]. The Hamiltonian matrix elements were obtained in the numerical localized orbital basis set implementation[32] within GGA. The atomic basis set (W7.0-s2p2d2f1 for Tungsten and Se7.0-s3p3d1 for Selenium) as well as other parameters were converged to a perfect agreement with reference GGA results of our Quantum-ESPRESSO calculations and Ref. 8.

**Acknowledgments:**

We thank Reyes Calvo for fruitful discussions. This research was supported by the VdW Heterostructure program (KCWF16) (STM spectroscopy and QPI mapping) funded by the Director, Office of Science, Office of Basic Energy Sciences, Materials Sciences and Engineering Division, of the US Department of Energy under Contract No. DE-AC02-05CH11231. Support was also provided by National Science Foundation award EFMA-1542741 (surface treatment and topographic characterization). The work at the ALS (sample growth and ARPES measurements) is supported by the Office of Basic Energy Sciences, US DOE under Contract No. DE-AC02-05CH11231. The work at the Stanford Institute for Materials and Energy Sciences and Stanford University (ARPES measurements) was supported by the Office of Basic Energy Sciences, US DOE under contract No. DE-AC02-76SF00515. S. T. acknowledges the support by CPSF-CAS Joint Foundation for Excellent Postdoctoral Fellows. H. R. acknowledges fellowship support from NRF, Korea through Max Planck Korea/POSTECH Research Initiatives No. 2016K1A4A4A01922028 and No. 2011-0031558. A.P. and O.V.Y. acknowledge support by the ERC Starting grant "TopoMat" (Grant No. 306504) (theoretical formalism development). Q.W. acknowledges support from NCCR Marvel (hybrid functional calculations). First-principles calculations were performed at the Swiss National Supercomputing Centre (CSCS) under project s675 and the facilities of Scientific IT and Application Support Center of EPFL. The work at Nanjing University (Y.Z.) is supported by the Fundamental Research Funds for the Central Universities No. 020414380037 (surface structure analysis). M.M.U. acknowledges support by Spanish MINECO under grant no. MAT2014-60996-R (data analysis).

**Author contributions:**

M.M.U., Y.Z. and S.K.M. conceived the work and designed the research strategy. M.M.U., A.M.R., Y.C., D.W. and Z.P. measured and analyzed the STM/STS data. Y.Z., H.Y. and S.T. performed the MBE growth and ARPES and XPS characterization of the samples. A.P. and Q.W. performed the theoretical calculations. F.W. participated in the interpretation of the experimental data. S.K.M., and Z.X.S. supervised the MBE growth and ARPES and XPS characterization. O.V.Y. supervised the calculations. M.F.C. supervised the STM/STS experiments. M.M.U. wrote the paper with help from O.V.Y. and M.F.C. M.M.U. and M.F.C. coordinated the collaboration. All authors contributed to the scientific discussion and manuscript revisions.

**Competing financial interests**

The authors declare no competing financial interests.

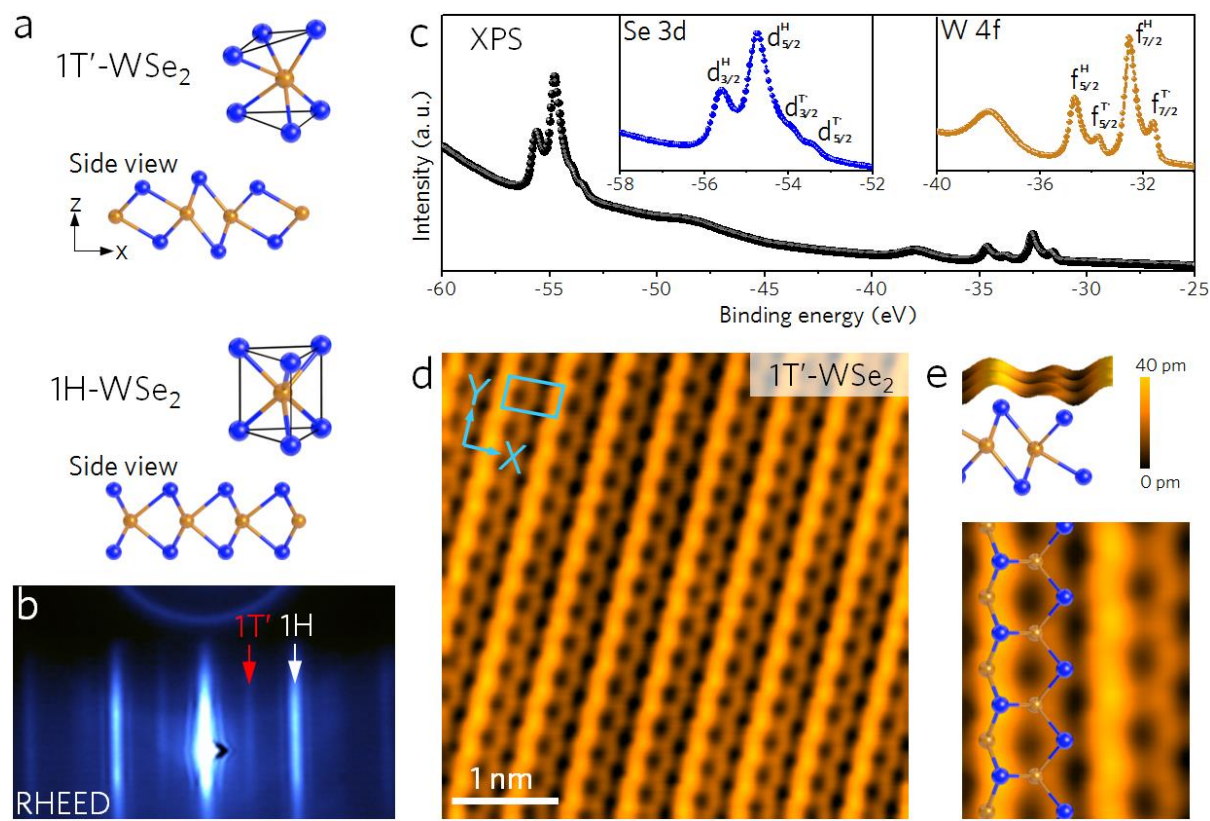

**Figure 1. Atomic structure of mixed-phase single-layer WSe$_2$. a**, Calculated unit cells and side view sketches of the 1T' and 1H phases of single-layer WSe$_2$. Se (W) atoms are depicted in blue (orange). **b**, RHEED pattern of single-layer 1T'/1H-mixed phase WSe$_2$. Red and white arrows indicate diffraction stripes from 1T' and 1H phases, respectively. **c**, Core level XPS spectrum of single-layer 1T'/1H mixed phase WSe$_2$. Insets show zoom-in of the Se (blue) and W (orange) peaks for the 1T' (d$^{T'}$, f$^{T'}$) and 1H (d$^H$, f$^H$) phases. **d**, Atomically-resolved STM image of single-layer 1T'-WSe$_2$. The unit cell is indicated in blue (V$_s$ = + 500 mV, I$_t$ = 1 nA). **e**, Side and top view close-up of 1T'-WSe$_2$ STM image with sketch of calculated 1T'-WSe$_2$ (only upper layer Se atoms are depicted in top view).

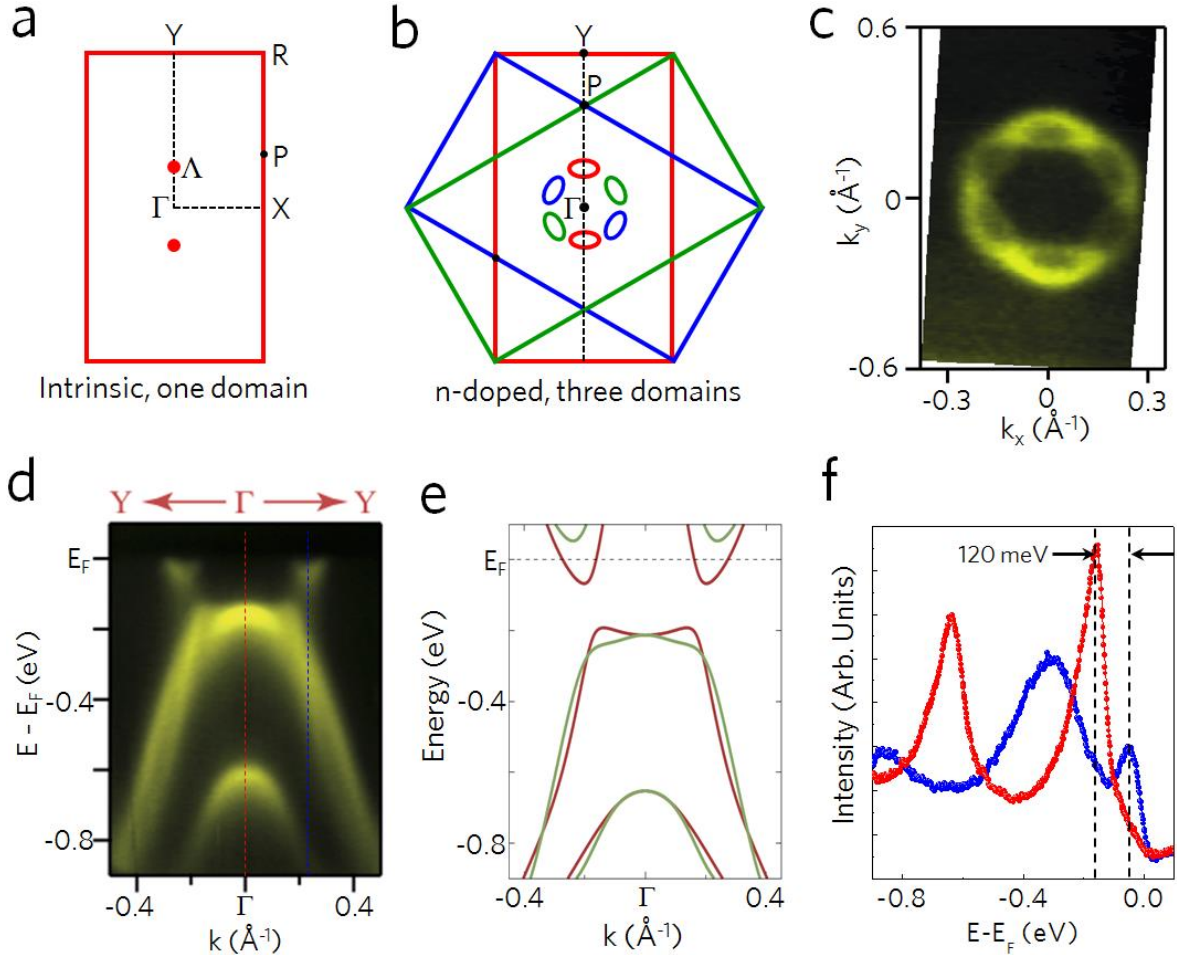

**Figure 2. ARPES characterization of single-layer 1T'-WSe$_2$. a**, Sketch of the first Brillouin zone of 1T'-WSe$_2$. Relevant high-symmetry points are indicated. **b**, Three surface Brillouin zones corresponding to the three rotational 1T'-WSe$_2$ domains on the BLG surface represented by three different colors. The Fermi surface pockets from each rotational domain are indicated by ellipses of corresponding colors. Black dashed line represents the experimental ARPES line cut shown in (**d**). **c**, Experimental 1T'-WSe$_2$ Fermi surface measured by ARPES. **d**, High resolution ARPES band dispersion along the Y-Γ-Y direction. Due to the presence of rotational domains, contributions from both Γ-Y and Γ-P directions are observed in a single ARPES measurement (T = 60 K and photon energy E = 75 eV). **e,** Calculated bands for the 1T' phase of single-layer WSe$_2$ along Γ-Y (brown) and Γ-P (green) directions. A downward rigid shift of 130 meV has been added to account for n-doping seen in the experiment. **f**, EDCs from the momentum positions marked with dashed blue and red lines in **d**.

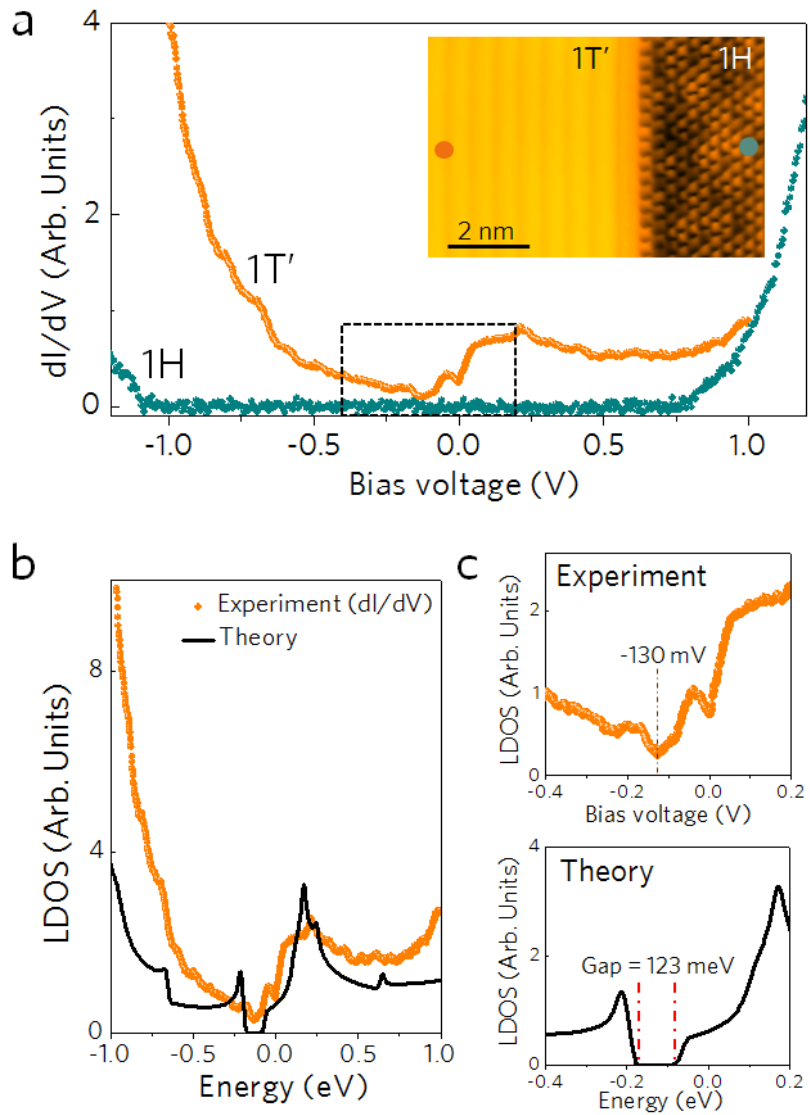

**Figure 3. STS characterization of single-layer mixed-phase $WSe_2$. a**, STS spectra obtained in the 1T' (orange) and 1H (blue) regions of single-layer $WSe_2$ (f = 614 Hz, $I_t$ = 0.3 nA, $V_{rms}$ = 4 meV). The inset shows an STM image of coexisting 1T' and 1H regions with a well-ordered interface between them ($V_s$ = + 500 mV, $I_t$ = 0.1 nA). **b**, Calculated LDOS(E) of bulk single-layer 1T'-$WSe_2$ (black curve) compared to experimental STS spectrum (orange curve). **c**, (Upper panel) Close-up view of the boxed region in **a** shows low-energy experimental STS spectrum taken for 1T'-$WSe_2$ phase. (Lower panel) Calculated LDOS(E) for 1T'-$WSe_2$ over the same energy range as upper panel.

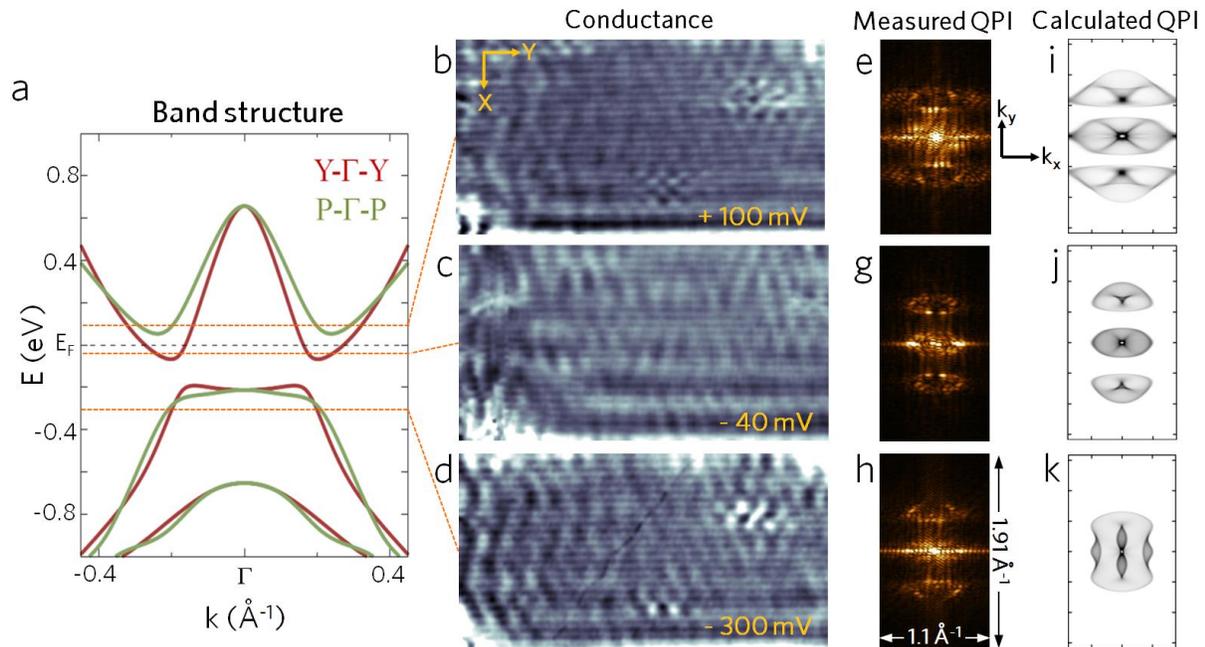

**Figure 4. Quasiparticle interference patterns in single-layer 1T'-WSe$_2$. a**, Calculated band structure of single-layer 1T'-WSe$_2$ along Γ-Y (brown) and Γ-P (green) directions in the ± 1 eV range. **b-d**, Experimental dI/dV conductance maps taken at (**b**) $V_s$ = + 100 mV, $I_t$ = 0.15 nA, (**c**) Vs = - 40 mV, $I_t$ = 0.15 nA, and (**d**) Vs = - 300 mV, $I_t$ = 0.15 nA (14 nm x 26.4 nm, f = 614 Hz, $V_{rms}$ = 4 meV). **e-h**, FFTs of the conductance maps in **b-d**. **i-k**, Calculated QPI patterns for (**i**) E = + 100 meV, (**j**) E = -40 meV, and (**k**) E = - 300 meV.

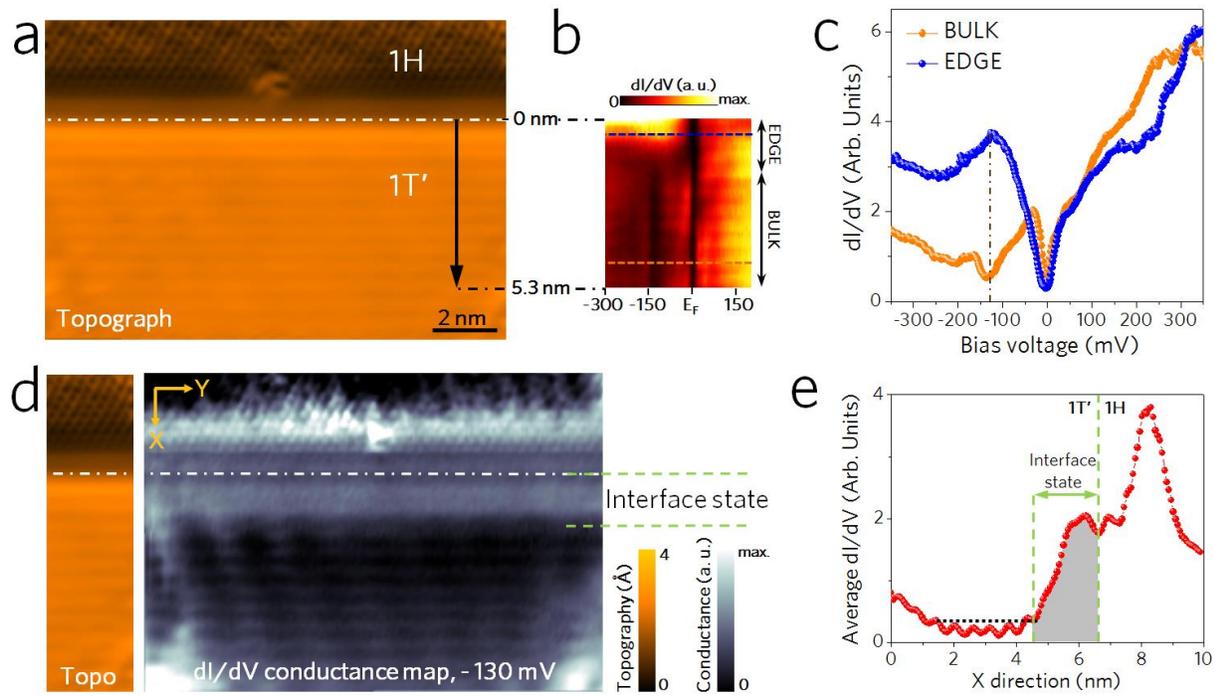

**Figure 5. Spatial extent of helical interface state in single-layer 1T'-WSe$_2$. a**, STM topograph of the 1T'-1H interface ($V_s$ = - 525 mV, $I_t$ = 0.2 nA). Dashed line shows interface location (see text). **b**, Color-coded dI/dV spectra taken along the path marked by the arrow in **a** (f = 614 Hz, It = 0.6 nA, $V_{rms}$ = 4 meV). **c**, dI/dV curves extracted from **b**. **d**, Experimental dI/dV map taken in the same region as **a** for $V_s$ = - 130 meV. Dashed line shows same interface location as in **a**. **e.** Average dI/dV linescan oriented along the X direction (vertical yellow arrow) for $V_S$ = -130 mV.

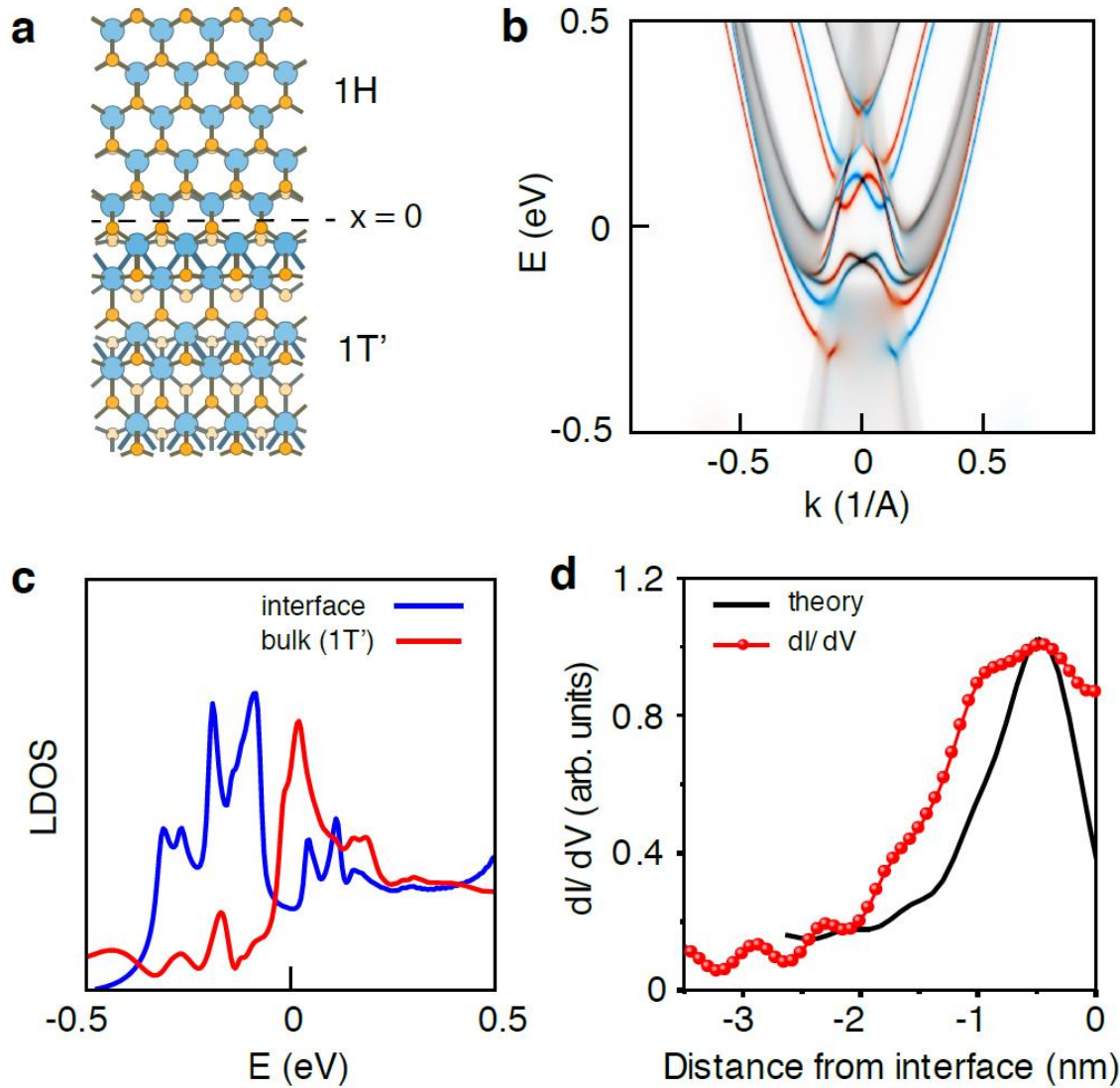

**Figure 6. WSe$_2$ 1T'-1H interface electronic structure. a,** Sketch of the structural model used to theoretically investigate the 1T'-1H interface in single-layer WSe$_2$. The interface position $x = 0$ is indicated. **b,** Momentum- and spin-resolved LDOS(E) at the 1T'-1H interface shows the dispersion and spin-momentum locking of the interface states (blue/red curves show different spin polarizations). **c,** Energy-resolved LDOS at the 1T'-1H interface (blue curve) in single-layer WSe$_2$ compared to the LDOS at a point further into the 1 T' bulk region (red curve). **d,** Dependence of LDOS at the band gap energy on distance from the 1T'-1H interface compared to experimental dI/dV linecut at $V_s = -130$ mV (from Fig. 5e).

Supplementary Materials for

# Observation of Topologically Protected Edge-states at Crystalline Phase Boundaries in Single-layer WSe$_2$

Miguel M. Ugeda[*], Artem Pulkin, Shujie Tang, Hyejin Ryu, Quansheng Wu, Yi Zhang, Dillon Wong, Zahra Pedramrazi, Ana Martín-Recio, Yi Chen, Feng Wang, Zhi-Xun Shen, Sung-Kwan Mo, Oleg V. Yazyev and Michael F. Crommie[*]

*Correspondence to: mmugeda@dipc.org and crommie@berkeley.edu.

# 1. Morphology of single-layer 1T'/1H mixed phase $WSe_2$

Figure S1 shows the typical morphology of our samples grown by holding the substrate temperature at 500 K, a temperature 175 K lower than that used to grow the 1H phase of $WSe_2$ (675 K). Under these growth conditions, the surface shows a similar morphology to other MBE-grown TMDs at higher temperatures[1,2]: Large regions of single layer TMD usually decorated with small islands of bilayers. However, ~ 15% of the total area of $WSe_2$ now exhibits the 1T' phase, which grows both laterally to the 1H phase and as islands in the first and second layer. The observation of a significant fraction of the 1T' phase is not unexpected given the relatively small energy difference between the stable 1H phase and metastable 1T' phase (0.3 eV per unit cell)[3]. The H and T' regions as well as the straight interfaces (green arrows) are indicted in fig. S1.

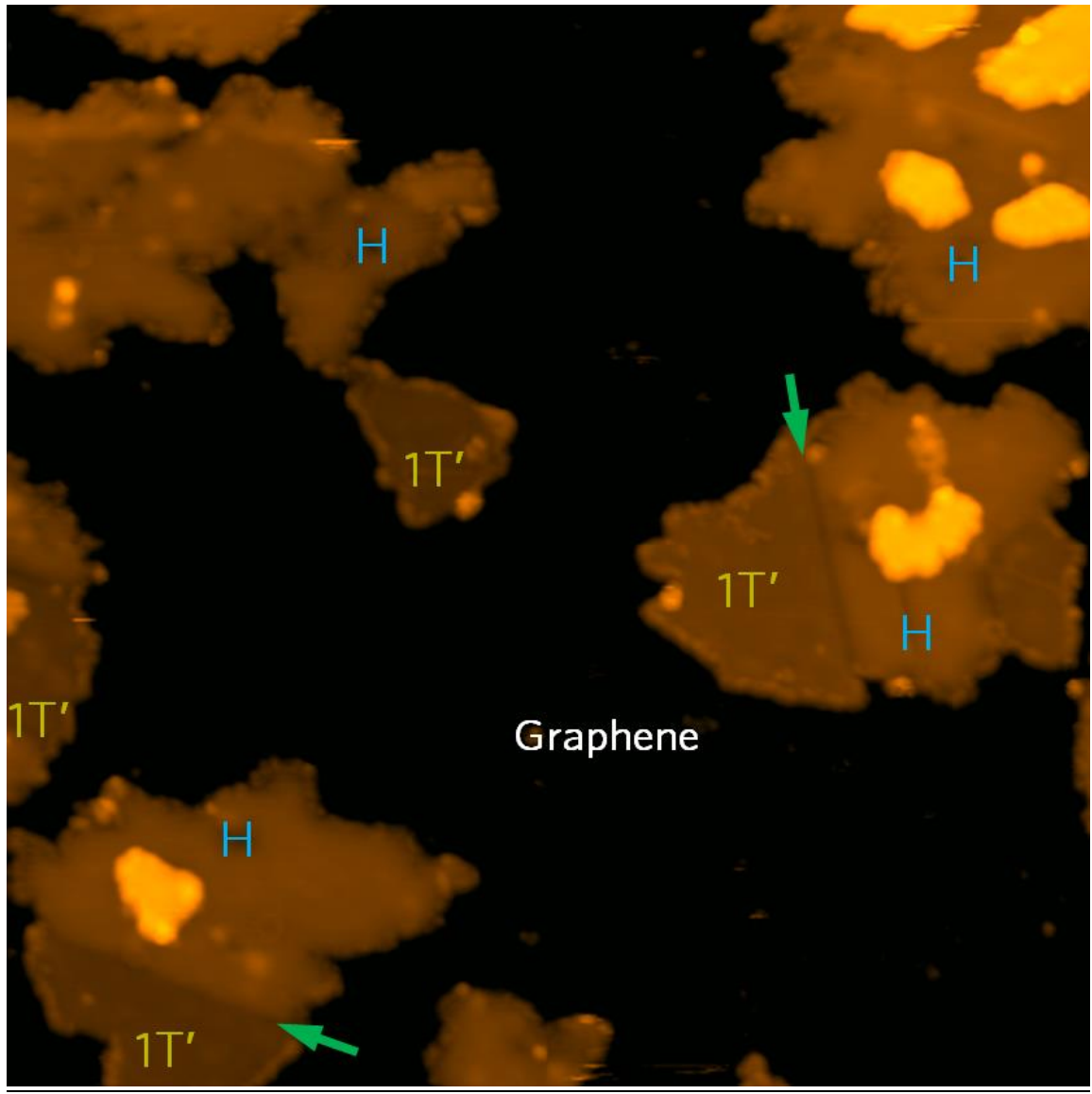

**Figure S1. STM characterization of single-layer 1T'/1H mixed phase WSe$_2$.** STM image shows large-scale view of the WSe$_2$/BLG samples studied in this work. Parameters: 1000 Å x 1000 Å, $V_s$ = + 1.8 V, $I_t$ = 10 pA, T = 5 K.

## 2. ARPES Fermi surface contour of multi-domain 1T'-WSe$_2$

Since we grow the two-fold symmetric 1T'-WSe$_2$ on top of the three-fold symmetric bilayer graphene (BLG), there naturally exist three energetically equivalent domains rotated by 120º with respect to each other. Such multiple-domain structure is commonly observed, particularly when a sample and substrate have different symmetries. A well-known example of this is Bi$_2$Se$_3$ on Bi$_2$Sr$_2$CaCu$_2$O$_{8+\delta}$ [4,5] Fig. S2 shows how multiple domain structure in 1T'-WSe$_2$/BLG affect the observed Fermi surface (FS) through ARPES. When there exists a single domain of 1T'-WSe$_2$, the expected FS is made of two small ellipses (black) near the Γ-point in a rectangular surface Brillouin zone (red). The second domain in the sample, rotated by 120º compared to the first one (green), causes new FS pockets (light green) to become superimposed over the contribution from the first domain. The same holds true for the third domain (blue) and corresponding FS pockets (blue). In the extended zone scheme (bottom panel), a complex FS is obtained (black) formed by elliptical contributions from all three domains. This is well observed in our experimental data presented in Fig. 2.

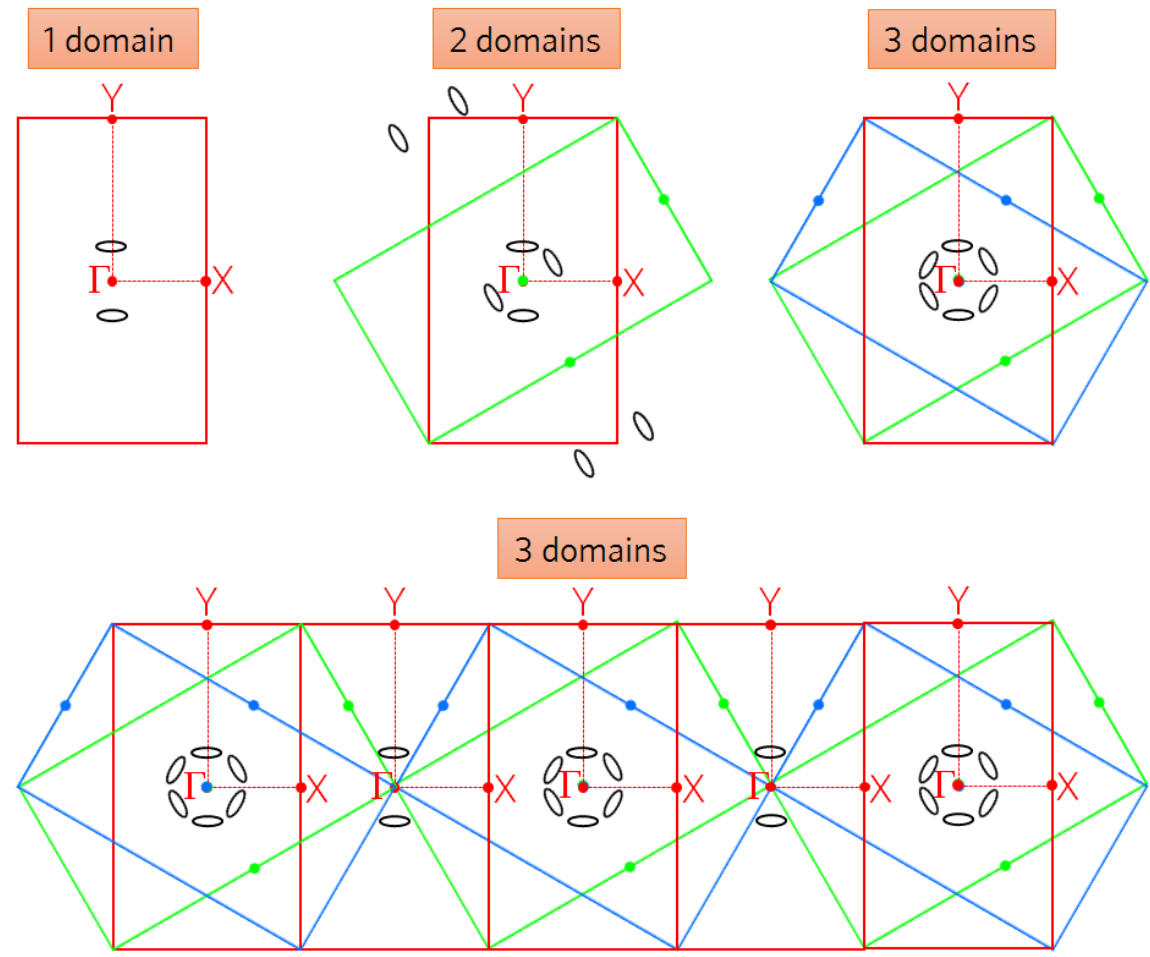

**Figure S2. Complex experimental Fermi surface topology due to the multi-domain structure of MBE grown 1T'-WSe2 on bilayer graphene.** The Brillouin zones corresponding to the three rotational domains are drawn in red, blue and green.

## 3. Single-layer 1T'-WSe$_2$ STS gap determination

The average STS gap value of 85 ± 21 mV was determined through statistical analysis of 99 d$I$/d$V$ spectra collected at the surface of monolayer 1T'-WSe$_2$ for numerous different 1T'-WSe$_2$ islands and numerous different STM tips. Fig. S3a shows the analysis procedure for a typical STS spectrum. We first identified the minimum of the gap (*Min*) and two peak features on both sides of the gap ($P_L$ and $P_R$). The full width between the two half-height points ((*Min*+ $P_L$)/2 and (*Min*+ $P_R$)/2) is used to define the STS gap width. Analysis of 99 d$I$/d$V$ spectra yields an average gap width of 85.1 mV with a standard deviation of 21 mV. A histogram of the STS gap width is shown in Fig. S3b.

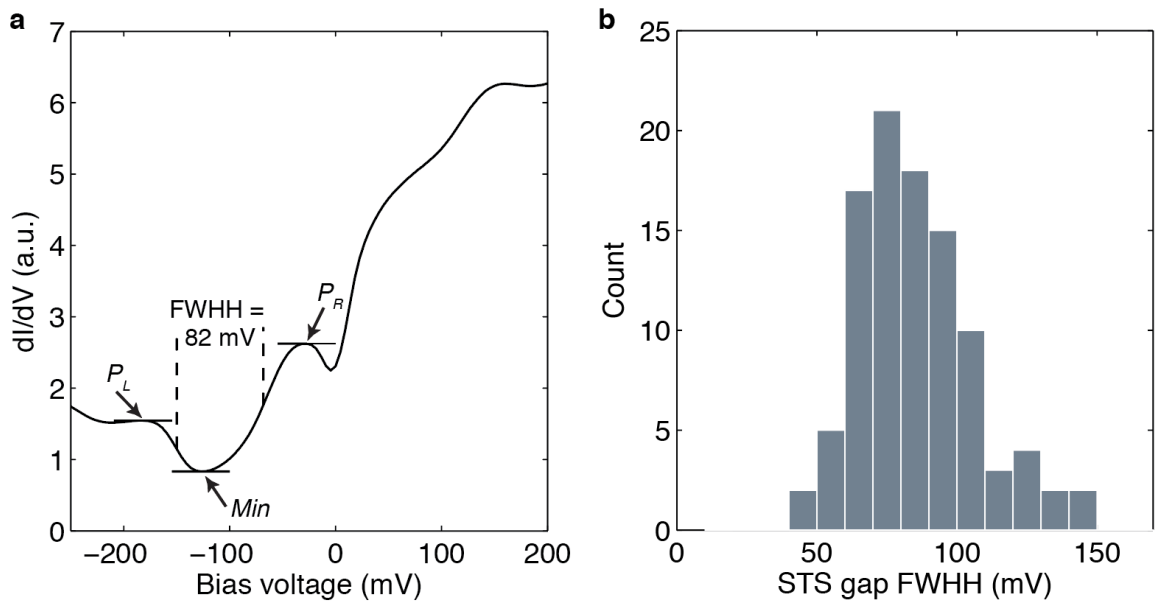

**Figure S3. STS gap determination. a,** Analysis of a typical d$I$/d$V$ spectrum. The minimum of the gap (*Min*) and peak features ($P_L$ and $P_R$) are identified. The full width between the two half-height points ((*Min*+ $P_L$)/2 and (*Min*+ $P_R$)/2) is used to define the STS gap width. **b,** Histogram of the STS gap width.

The STS gap width defined this way (85 ± 21 meV) is smaller than the measured ARPES gap (120 ± 20 meV) and the band gap calculated using the hybrid functional DFT approach (123 meV). The observed STS spectra also finite spectral weight inside the gap rather than showing a full gap as expected from the DFT calculations. This can be explained by lifetime broadening in the STS spectra of single-layer 1T'-WSe$_2$ as shown in Fig. S4.

Figure S4 compares the simulated lifetime-broadened LDOS curves (Figs. S4b-g) to a typical experimental d$I$/d$V$ spectrum (Fig. S2h). The lifetime-broadened LDOS curves were calculated by convoluting the LDOS spectrum (Fig. S3a) with a Lorentzian function having a broadening parameter $\Gamma$[6]

$$\text{LDOS}(E,\Gamma) = \int dE' \frac{\Gamma}{(E-E')^2 + \Gamma^2} \times \text{LDOS}(E',0)$$

As $\Gamma$ is increased, the spectral weight inside the gap fills up. Even with a small broadening of $\Gamma$ = 10 meV, the gap is no longer full. A FWHH analysis of the broadened LDOS curves shows that the effective gap size decreases with increasing $\Gamma$. When $\Gamma$ is 30 meV, the overall shape of the broadened LDOS curve strongly resembles the experimental d$I$/d$V$ spectrum, implying the significance of lifetime broadening in the collected STS spectra.

It is worth noting that lifetime effects should also broaden the ARPES spectra. An estimation of $\Gamma$ = 30 meV corresponds to FWHH = 60 meV, which is comparable to the FWHH of the ARPES EDC at gap edges in Fig. 2f. However, extracting the exact quasiparticle lifetime from the ARPES spectra requires a detailed understanding of all sources of inhomogeneous broadening (e.g. defects), which we presently do not have.

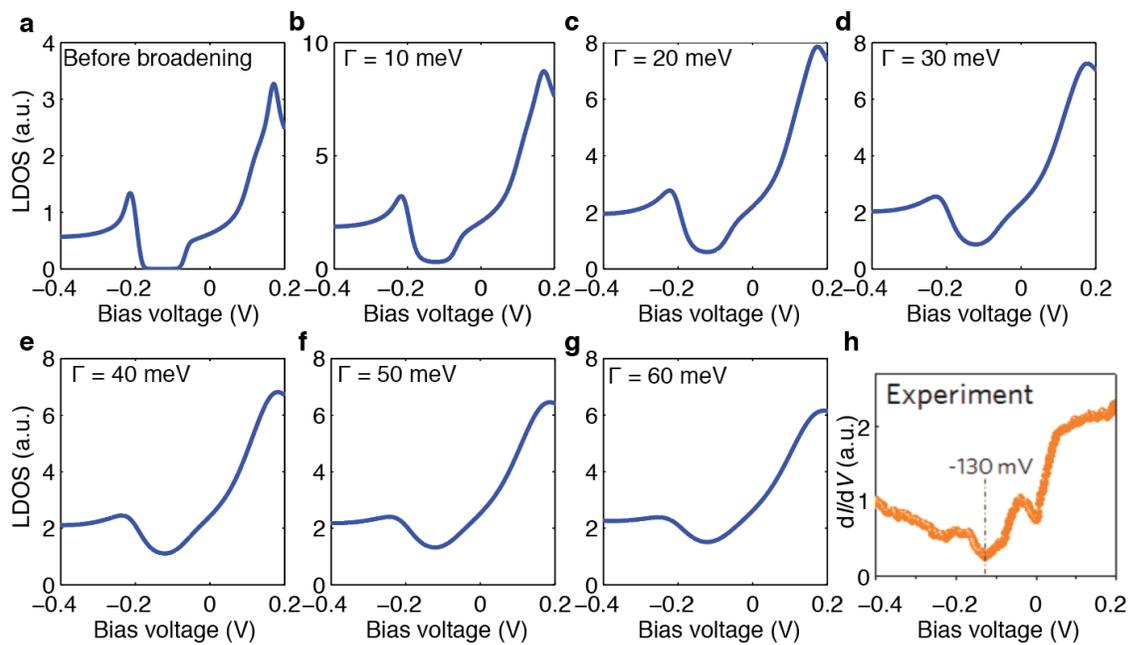

**Figure S4. Comparison between lifetime-broadened LDOS curves and an experimental STS spectrum. a,** Theoretical LDOS spectra before broadening. **b-g**, Broadened LDOS curves with different broadening parameters $\Gamma$. **h,** Experimental d$I$/d$V$ spectrum.

## 4. Observation of topologically protected edge states at irregular edges

The topological protection of boundary states in a QSHI phase implies that they should survive regardless of the structural features of the boundary, including different levels of interface roughness and defects, so long as time-reversal symmetry is preserved. In order to test this prediction in single-layer 1T'-WSe$_2$ we performed STS measurements of disordered edges of 1T'-WSe$_2$ islands as shown in Fig. S5a. Fig. S5b shows a color-coded plot of a series of dI/dV curves taken along the black arrow in Fig. S5a, which demonstrate a transition in electronic behavior from bulk to edge. Similar to the 1T'/1H interface, the gap feature at −130 meV abruptly disappears in a narrow strip near the edge and a peak emerges in the LDOS at the bulk gap energy. Here, however, the width of the edge state cannot be accurately measured due to disorder in the structure of the 1T'-vacuum interface. The fact that we observe the edge states regardless of edge structural details provides further evidence that we are observing a topologically-protected edge state rather than a trivial one.

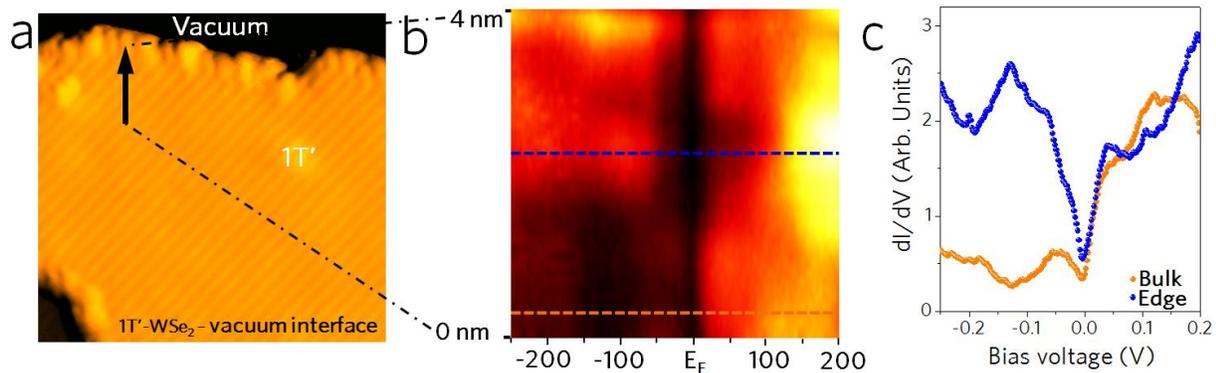

**Figure S5. Helical edge-state at disordered edges. a**, STM image of a 1T'-WSe$_2$ monolayer with irregular edges (17 nm x 16 nm, $V_s$ = + 1000 mV, $I_t$ = 0.01 nA). **b**, Color-coded dI/dV spectra taken along the path marked by the black arrow in **a** (f = 614 Hz, It = 0.1 nA, $V_{rms}$ = 4 meV). **c**, dI/dV curves extracted from **b**.